\documentclass[prl,superscriptaddress,amsmath,amssymb,showpacs,noshowkeys,a4paper,twocolumn]{revtex4}

\usepackage{graphicx}% Include figure files
\usepackage{dcolumn}% Align table columns on decimal point
\usepackage{bm}% bold math
\usepackage[usenames]{color}
\usepackage{epstopdf}
\usepackage{wasysym}
\definecolor{mygray}{gray}{0.5}

\newcommand{\affpof}{\affiliation{Physics of Fluids Group, MESA+ Institute for Nanotechnology, J. M. Burgers Centre for Fluid Dynamics, University of Twente, P. O. Box 217, 7500 AE Enschede, The Netherlands}}

\begin{document}

\title{How geometry determines the coalescence of low-viscosity drops}

\author{A.~Eddi}
\affpof
\author{K.G.~Winkels}
\affpof

\author{J.H.~Snoeijer}
\affpof

\begin{abstract}
The coalescence of water drops on a substrate is studied experimentally. We focus on the rapid growth of the bridge connecting the two drops, which very quickly after contact ensues from a balance of surface tension and liquid inertia. For drops with contact angles below $90^\circ$, we find that the bridge grows with a self-similar dynamics that is characterized by a height $h\sim t^{2/3}$. By contrast, the geometry of coalescence changes dramatically for contact angles at $90^\circ$, for which we observe $h\sim t^{1/2}$, just as for freely suspended spherical drops in the inertial regime. We present a geometric model that quantitatively captures the transition from $2/3$ to $1/2$ exponent, and unifies the inertial coalescence of sessile drops and freely suspended drops.
\end{abstract}

\maketitle
The splitting and merging of liquid drops are key processes during cloud formation, condensation and splashing \cite{Grabowski:2012,Andrieu:2002,Blaschke:2012}. The rate at which these processes take place is very important for technologies involving sprays and printing \cite{Eggers:2008,Wijshoff:2010}. Breakup and coalescence are singular events during which the liquid topology changes from a single drop to multiple drops, or vice versa \cite{Eggers:1997}. 
Near the singularity, i.e. right before the moment of pinch-off or just after coalescence has been initiated, a tiny bridge of liquid connects two macroscopic drops. The size of this bridge vanishes at the singularity and gives rise to power-law divergence of stress \cite{Eggers:1997}. In most cases, the dynamics near coalescence and pinch-off is universal in the sense that it is completely independent of initial conditions. In this regime viscosity, surface tension and inertia are all relevant \cite{Eggers:1997, Case:2008, Paulsen:2011, Paulsen:2012}.

For low-viscosity liquids such as water, however, most of the dynamics can be described by a fully inertial regime, where viscous forces can be neglected. For pinch-off, one observes that the bridge size vanishes as $\tau^{2/3}$ in this inertial regime \cite{Chen:1997,Day:1998,Chen:2002,Castrejon:2012}, where $\tau$ is the time to pinch-off. Interestingly, the situation is markedly different for the inertial coalescence of spherical water drops: the bridge grows with time as $t^{1/2}$ and the prefactor depends explicitly on the drop size \cite{Eggers:1999,Eggers:2003,Aarts:2005,Paulsen:2011, Paulsen:2012}. The outer scale enters the coalescence problem through the peculiar initial condition, which consists of two spheres touching at a single point. This geometry even modifies the scaling law with respect to pinch-off, as it strongly enhances the capillary forces that drive the coalescence.

\begin{figure}[t]
\begin{centering}
\includegraphics[width=.96\linewidth]{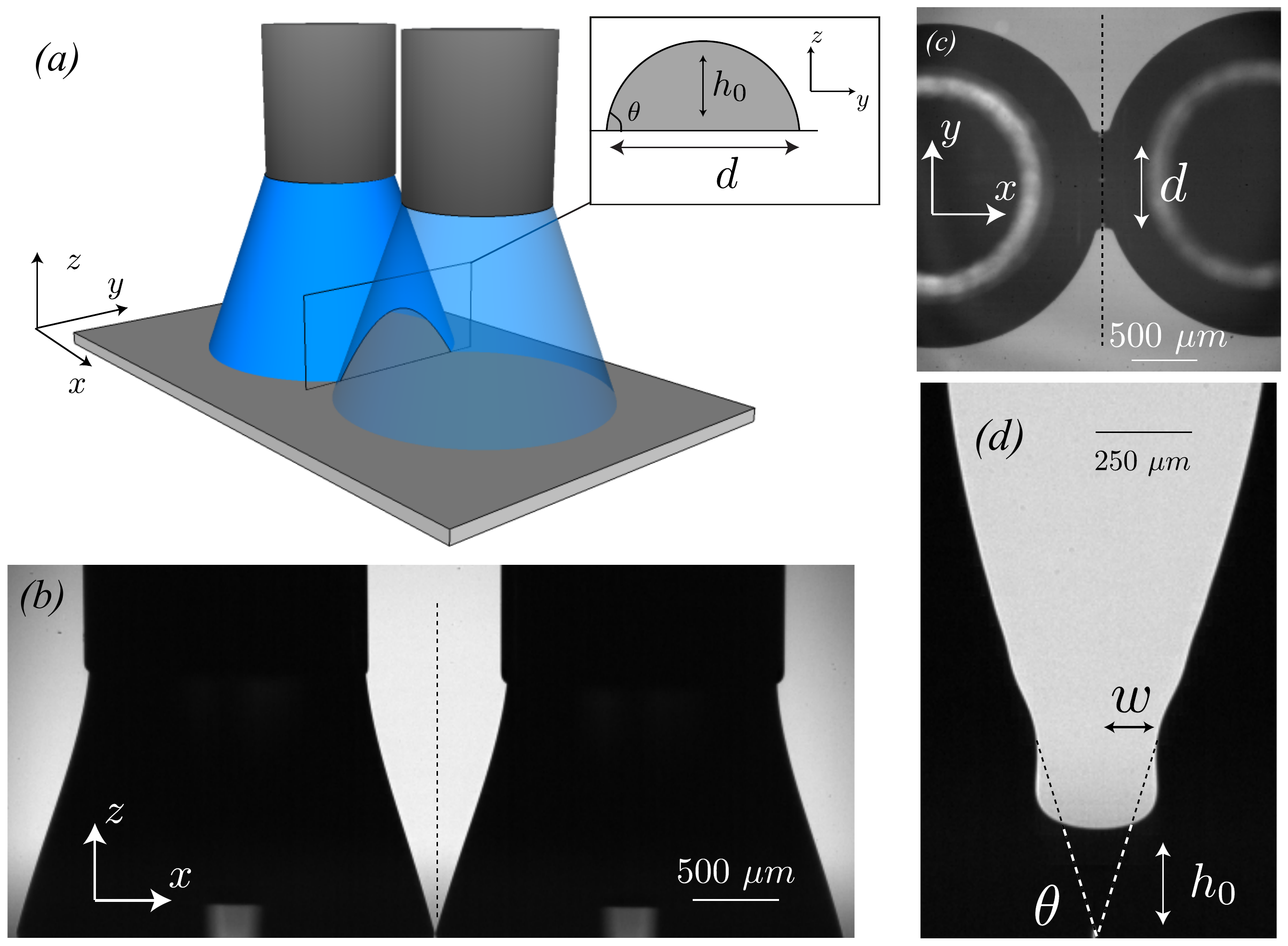}
\caption{\label{fig1} (a) Coalescence geometry for ``conical'' drops. The inset shows a cross-section of the liquid bridge, which here is sketched as a circle of contact angle $\theta$. (b) Side-view snapshot of the two drops on the last frame before contact. The two needles holding the drops can be seen at the top of the image. (c) Bottom-view image of the liquid bridge of width $d$ $500~\mu$s after contact. (d) Sideview of the bridge of height $h_0$ that joins the two drops $278~\mu$s after contact. The advancing contact angle $\theta$ determines the geometry at contact.}
\end{centering}
\end{figure}

In this Letter we reveal the inertial coalescence dynamics of water drops on a substrate. This is of comparable practical interest to that of freely suspended drops, but now the influence of geometry becomes even more apparent (Fig.~\ref{fig1}): the drop shape looks very different when viewed from the side or from the top. This situation was recently investigated for high-viscosity drops \cite{Ristenpart:2006,Narhe:2008,Yarin:2012}, with evidence for self-similar dynamics during the initial growth \cite{Hernandez:2012}. For inertial sessile drops, by contrast, the coalescence dynamics has remained unknown. Two-dimensional inviscid theory for ``coalescing wedges'' suggest a 2/3 exponent \cite{Keller:1983,Keller:2000,Billingham:2005}, as for pinch-off, but experiments for water drops on a substrate were interpreted using the classical 1/2 law \cite{Kapur:2007}. In addition, it is not known whether and how the substrate wettability influences the coalescence. 

Here we access the coalescence of water drops on a substrate with previously unexplored length and time scales. We show that the coalescence dynamics displays self-similarity and we present scaling arguments that quantitatively account for all observations. The key result is that for $\theta < 90^\circ$ the inertial coalescence is self-similar and displays a 2/3 exponent. However, the range over which this asymptotics can be observed is continuously reduced upon approaching $\theta=90^\circ$. In this limit, our theory  and experiments recover the 1/2 exponent and thus unifies the coalescence of sessile drops and freely suspended drops.

\paragraph{Experimental setup.~---}

Two drops of pure water (milli-Q, surface tension $\gamma=72~$mN.m$^{-1}$) are grown at the tip of flat dispense needles and get in contact with transparent glass substrates that are covered with different coatings with a typical hysteresis of $10^\circ$, leading to (static) advancing contact angles $\theta$ ranging from $\theta=73^{\circ}$ to $90^{\circ}$. Coalescence is achieved by very slowly advancing the contact lines, the static advancing angle $\theta$ actually determines the interface angle at the moment of contact. It is measured for each experiment, with an error smaller than $2^\circ$. In order to avoid dynamical effects, the approach speed of the two contact lines is always smaller than $20~\mu$m.s$^{-1}$ (the initial speed of coalescence measured experimentally is about $2.5~$m.s$^{-1}$). The shape of each drop can be adjusted by moving the needle up and down. There is a position of the needle where the drop takes a nearly perfectly conical shape [Fig.~\ref{fig1}(a)]. Such conical shapes appear as wedges in side-view [Fig.~\ref{fig1}(b)], and are very advantageous for revealing coalescence dynamics -- as will be demonstrated, the straight interface enhances the range over which scaling can be observed. 

The spatial and temporal resolution required for capturing the fast stages of coalescence is achieved by combining high-speed recording and long-range microscopy. We use two synchronized high-speed cameras to image the coalescence simultaneously from the side and from below [Fig.~\ref{fig1}(b-c)]. A Photron SA1.1 is coupled to a long-distance microscope (Navitar 12X Zoom coupled with a 2X adapter tube) and records images from the side with a resolving power of 4.8~$\mu$m/pixel. Combined with backlight diffusive illumination, it can record up to 200.000 frames/s. For the bottom view imaging, we use an APX-RS combined with a Navitar 6000 lens with a resolution of 2.7~$\mu$m/pixel. Reflective illumination allows for a frame rate of 90.000 frames/s. At these time and length scales, all our measurements are in a purely inertial regime where the viscosity of water can be neglected \footnote{Viscous effects are relevant when the width of the bridge [$w$, defined in Fig.~\ref{fig1}(d)], is smaller than the viscous length $\ell_v=\mu^2/(\rho \gamma)$ \cite{Paulsen:2011, Paulsen:2012}, which for water is of the order of 10 nm. This cross-over occurs at times $t\sim1-100~$ns.}. Once the drops are in contact and coalesce we measure the shape and evolution of the liquid bridge using a custom-made edge-detection algorithm in Matlab. Contact time is chosen half-way in between the last frame where no bridge is observed and the first frame where we can measure the bridge height.

\begin{figure*}[t]
\begin{centering}
\includegraphics[width=.96\textwidth]{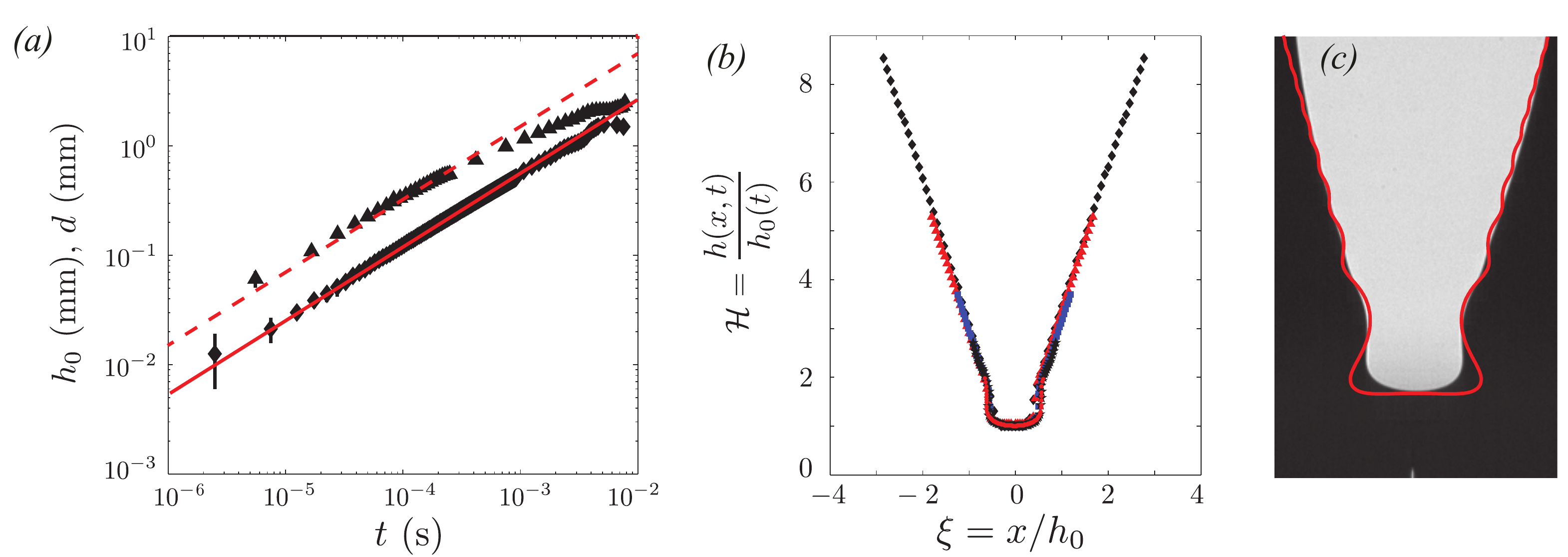}
\caption{\label{fig2} Coalescence of conical drops. (a) Height of the bridge $h_0$ (diamonds) and its width $d$ (triangles) as a function of time. Data are averaged over 6 experiments, statistical error bars indicate reproducibility. The contact angle is $\theta=73^{\circ}$ at the moment the drops come into contact. The solid line is the prediction~(\ref{eq:scalinglaw}) with a prefactor $D_0=0.89$. The dashed corresponds to~(\ref{eq:conversionbottomside}). (b) Rescaled profiles ${\cal H}=h(x,t)/h_0(t)$ versus $\xi=x/h_0(t)$ for 5 different times $ t=27.5, 72.5, 122.5, 197.5~$and $ 397.5~\mu$s after contact. The collapse reveals self-similar dynamics of our experimental measurements. (c) Comparison of the experimental profile (background snapshot) and the numerical similarity solution from two-dimensional potential flow (solid line, from \cite{Keller:2000,Billingham:2005}). The numerical curve actually corresponds to $\theta=75^\circ$, which was slightly rescaled here to match $\theta=73^\circ$.}
\end{centering}
\end{figure*}

\paragraph{Conical drops.~---}

A snapshot of the bridge profile during coalescence of two conical drops is presented in Fig.~\ref{fig1}(d). In this side-view, the drops appear as nearly perfect wedges of contact angle $\theta=73^\circ$, connected by a thin bridge of height $h_0$. The bridge height grows rapidly in time and emits a capillary wave on the surface of the drops, as can be seen from the still image. Note that the bridge shape is markedly different from highly viscous coalescing drops, for which no such waves were observed~\cite{Hernandez:2012}. This suggests that for water drops, which have a low viscosity, the dynamics is limited by liquid inertia rather than by viscous effects. 

To reveal the growth dynamics of the bridge, we measure the evolution of the minimum height $h_0$ as a function of time after contact $t$ [Fig.~\ref{fig2}(a)]. Experiments are extremely reproducible and data correspond to an average over 6 different coalescence events. The diamonds in Fig.~\ref{fig2}(a) clearly suggests a power-law growth of the minimum height, $h_0 \sim t^{2/3}$. This regime over which this scaling is observed covers more than 3 decades in time and more than 2 in space, until $h_0$ is of the order of the initial size of the conical drops (around $1$ mm) at time $t>5~$ms. Inspired by results from pinch-off \cite{Eggers:1997}, we now attempt to collapse the experimental bridge profiles using a similarity Ansatz. We therefore suggest that the bridge dynamics is governed by a similarity solution of the form
\begin{equation}
h(x,t)=h_0(t) {\cal H}(\xi), \quad {\rm with} \quad \xi = x/h_0.
\end{equation}
This scaling is verified in Fig.~\ref{fig2}(b), comparing side-view profiles for a single experiment at 5 different times after contact. The collapse of the experimental data confirms that the bridge shape is preserved during the inertial stages of coalescence. This self-similarity implies that the bridge height $h_0$ is the only relevant scale, and that the needle size is not important for the dynamics as viewed from the side. In particular, it means that the width of the bridge, defined in Fig. \ref{fig1}(d), scales as $w \sim h_0$.

A $2/3$ exponent  was previously observed for inviscid liquids, both for pinch-off of drops \cite{Day:1998,Chen:2002} and for merging of two-dimensional wedges \cite{Keller:1983,Keller:2000,Billingham:2005}, originating from a balance between surface tension, $\gamma$, and inertia characterized by the liquid density, $\rho$. Here we adapt this scaling law for the case of drop coalescence on a substrate. The driving pressure for $\theta$ close to (but smaller than) 90$^\circ$ is given by
%In the present context, however, the force balance has singular behavior when the contact angle approaches $\theta = 90^\circ$. This can be seen from the capillary pressure in the bridge, which scales as
%
\begin{equation}
%P_{cap}\sim \frac{\gamma}{w}~~\textrm{with}~~w=\frac{h_0}{\tan{\theta}},
P_{cap}\sim \frac{\gamma}{w}, \quad \textrm{with} \quad w=\left(\frac{\pi}{2} - \theta \right)h_0.
\label{eq:wedgeshape} 
\end{equation}
The width $w$ is defined in Fig.~\ref{fig1}(d) and provides the characteristic curvature $\kappa_{xz} \sim 1/w$ of the interface in the $(x,z)$-plane. The curvature in the $(y,z)$-plane $\kappa_{yz}$ is of opposite sign, but must be smaller in absolute magnitude to induce coalescence. The curvature diverges at the moment of contact where the bridge size goes to zero. The very strong capillary pressure (\ref{eq:wedgeshape}) drives the rapid flow of liquid into the bridge, yielding a dynamical pressure
\begin{equation}
P_{iner}\sim \rho v^2\sim \rho \left({\frac{h_0}{t}}\right)^2.
\end{equation}
Balancing these two pressures, one gets the scaling law for the bridge growth
\begin{equation}
%h_0=D_0 \Big(\frac{\gamma \tan{\theta}}{\rho}\Big)^{1/3}  t^{2/3}, 
h_0=D_0 \left[ \frac{\gamma}{\rho \left(\frac{\pi}{2}-\theta\right)}\right]^{1/3}  t^{2/3}.
\label{eq:scalinglaw}
\end{equation}
Fitting this to the diamonds in Fig.~\ref{fig2}(a) gives a numerical constant of order unity, $D_0=0.89$, suggesting that the balance is correct. This scaling law and the underlying similarity hypothesis must break down when the contact angle $\theta \rightarrow 90^\circ$, due to the vanishing denominator of (\ref{eq:scalinglaw}). 

Before pursuing this limit in more detail, it is interesting to compare our experiments to previously obtained similarity solutions for two-dimensional coalescing wedges \cite{Keller:2000,Billingham:2005}. While the scaling law in this two-dimensional inviscid theory is consistent with (\ref{eq:scalinglaw}), the numerically obtained profile $\cal{H}(\xi)$ does not capture the shape observed experimentally. This is revealed in Fig.~\ref{fig2}(c), where we overlap the experimental profile and the two-dimensional potential flow solution (red solid line). Unlike the experiment, the theoretical profile is extremely flattened at the bottom of the bridge, and the capillary waves on the drop surface are much more pronounced. This discrepancy with two-dimensional theory suggests that the three-dimensional nature of coalescence cannot be ignored. Namely, the bridge exhibits a second curvature in the $(y,z)$-plane, which scales as $\kappa_{yz} \sim h_0/d^2$ [Fig. \ref{fig1}(a), inset]. Since the bridge topology is a ``saddle'', this curvature is of opposite sign compared to $\kappa_{xz} \sim 1/w$. The curvature $\kappa_{yz}$ can influence the coalescence, provided that it is of comparable magnitude -- the width of the bridge $d$ should thus have the same 2/3 power-law evolution during the inertial growth. Assuming the cross-section in Fig.~\ref{fig1}(a) is a circular arc of (advancing) contact angle $\theta$, one actually predicts

\begin{equation}
\frac{d}{h_0}=\frac{2 \sin{\theta}}{1-\cos{\theta}}.
\label{eq:conversionbottomside}
\end{equation}
To test this hypothesis we measure $d$ from the bottom-view [Fig.~\ref{fig2}(a), triangles]. Indeed, the measured dynamics for $d(t)$ is consistent with the $2/3$ power law. The dashed line is the prediction by (\ref{eq:conversionbottomside}) with $\theta=73^\circ$. This static advancing contact angle indeed gives a good description of the data; in line with earlier findings for viscous drops on a substrate \cite{Ristenpart:2006,Narhe:2008,Hernandez:2012}, this suggests that contact line motion is not the rate-limiting factor for the bridge growth. 
We thus conclude that both principal curvatures are comparable $|\kappa_{xz}| \sim |\kappa_{yz}|$, though from (5) the latter is indeed of smaller magnitude. This makes the problem inherently three-dimensional (Fig.~\ref{fig1}). 

\paragraph{Spherical drops.~---}

\begin{figure*}[t]
\begin{centering}
\includegraphics[width=.96\linewidth]{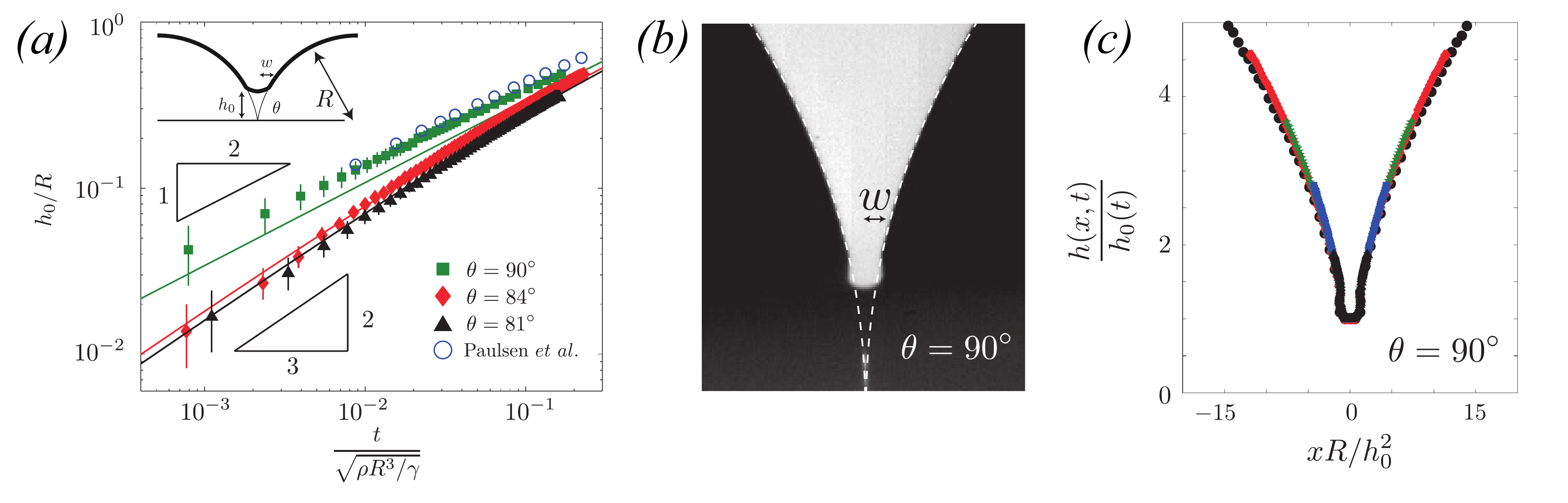}
\caption{\label{fig3}  (a) Growth of the bridge height $h_0$ as a function of time for three different contact angles $\theta=81^{\circ}$ (triangles, $R=1.8~$mm), $\theta=84^{\circ}$ (diamonds, $R=1.5~$mm), and $\theta=90^{\circ}$ (squares, $R=1.9~$mm). The open circles are experimental data points from Paulsen \textit{et al.} \cite{Paulsen:2012}, for freely suspended water drops. The continuous lines are predictions from (\ref{eq:finitesizeeffects}), with $D_0=0.89$. The inset shows the geometry for spherical drops, and defines the meniscus size $w$. (b) Snapshot of the growing bridge for $\theta=90^{\circ}$. (c) Rescaled profiles $h(x,t)/h_0(t)$ versus $xR/h_0^2(t)$ for $\theta=90^{\circ}$ and $ t=25, 75, 125, 225~$and $ 550~\mu$s after contact.}
\end{centering}
\end{figure*}

A more natural geometry for drop coalescence is of course encountered for spherically shaped drops  deposited on a substrate with a contact angle $\theta$ (Fig.~\ref{fig3}). A very special case is obtained for $\theta=90^\circ$: if the substrate is considered as a ``mirror", the geometry is identical to that of two freely suspended spherical drops. To explore this special case we have performed experiments with $\theta=90^\circ$ [Fig.~\ref{fig3}(b)]. The result is presented in Fig. \ref{fig3}(a) (green squares), clearly showing that $h_0 \sim t^{1/2}$. On the same graph we replot measurements for freely suspended spherical water drops in the inertial regime  (blue open circles, taken from \cite{Paulsen:2012}): within experimental error, the coalescence of free water drops is indeed identical to that of drops on a substrate of $\theta=90^\circ$. This suggests that our experiments can be interpreted using the same argument as for freely suspended drops. The key difference with respect to Eq.~(\ref{eq:wedgeshape}) is that the driving radius of curvature now reads $w \sim h_0^2/R$ -- the pressure balance then leads to the 1/2 exponent \cite{Eggers:1999,Aarts:2005,Paulsen:2012}. To further test this argument, we once more attempt a similarity Ansatz, but now rescaling the horizontal coordinate as $x/w 
\sim xR/h_0^2$ instead of $x/h_0$. Figure~\ref{fig3}(c) indeed gives a collapse and confirms this scaling. 

Intriguingly, we thus find that the coalescence exponent changes from $2/3$ obtained with $\theta=73^\circ$, to $1/2$ for $\theta=90^\circ$. To investigate the transition, we performed experiments using sessile droplets with contact angles close to but lower than $90^{\circ}$. Results for $\theta=81^\circ$ and $\theta=84^\circ$ are presented on Fig. \ref{fig3}(a) (black triangles, red diamonds). Both data sets clearly show $h_0\sim t^{2/3}$. We now develop a geometric model that explains the change in exponents near $90^{\circ}$. The model considers the geometry sketched in the upper inset of Fig. \ref{fig3}(a), consisting of two intersecting circles of radius $R$, from which we find the meniscus size:
\begin{equation}
\frac{w}{R}=\sin{\theta}-\left[1-\left( \frac{h_0}{R}+\cos{\theta}  \right)^2  \right]^{1/2}.
\label{eq:sphericalshape}
\end{equation}
While for asymptotically small $h_0$ and $\theta < 90^\circ$ this expression is identical to the wedge-shape of (\ref{eq:wedgeshape}), $w\sim h_0$, it also captures the spherical geometry at $\theta=90^\circ$ where the wedge disappears and $w \sim h_0^2/R$. With this refinement, the bridge dynamics during the inertial regime can be predicted for all contact angles
 
\begin{equation}
\frac{D_0^3\gamma t^2}{\rho} = h_0^2 R \left[\sin{\theta}-\left[1-\left( \frac{h_0}{R}+\cos{\theta}  \right)^2  \right]^{1/2}\right].
\label{eq:finitesizeeffects}
\end{equation}
The results of this theoretical prediction are shown on Fig. \ref{fig3}(a) with no adjustable parameter (we keep $D_0=0.89$ as obtained from Fig.~\ref{fig2}). The theory indeed captures the experimentally observed $\theta$ dependence. 

The sudden transition of exponent from $2/3$ to $1/2$ at $\theta=90^\circ$ can now be understood as follows. Taking first the angle $\theta=90^\circ$ and then the early-time limit $t \rightarrow 0$, Eq.~(\ref{eq:finitesizeeffects}) gives an exponent 1/2. On the other hand, taking first $t \rightarrow 0$ at $\theta < 90^\circ$ and then the limit $\theta \rightarrow 90^\circ$ gives 2/3. This is not inconsistent, since the range over which the 2/3-asymptotics applies vanishes near $90^\circ$, as $h_0/R \ll (\pi/2 - \theta)$: the duration of the 2/3 regime gradually shrinks to zero. Yet, for $\theta$ only a few degrees smaller than 90$^\circ$, the 2/3 exponent can still be observed [cf. Fig.~\ref{fig3}(a)]. 

%Indeed, this expression is equivalent to (\ref{eq:scalinglaw}) for $\theta<90^{\circ}$ and $h_0\rightarrow 0$ and leads to $h_0\sim t^{2/3}$. From (\ref{eq:finitesizeeffects}), this asymptotics is valid over a limited range $h_0 \ll R (\pi/2 - \theta)$, for which the unperturbed meniscus can be approximated by a wedge. Upon approaching $90^{\circ}$, the duration of this $2/3$ regime thus shrinks continuously until it vanishes at $90^{\circ}$; in this limit one recovers recovers $h_0\sim t^{1/2}$, in full analogy to freely suspended drops in the inertial regime \cite{Eggers:2003,Aarts:2005, Paulsen:2012}. Keeping the prefactor fixed at $D_0=0.89$, as is shown in Fig. 3, 
 
\paragraph{Discussion.~---}
Our results reveal that coalescence dynamics in the inertial regime is dictated by the geometry at the moment two drops come into contact. 
%For contact angles well below $90^\circ$, the initial geometry is equivalent to that of two wedges. In that case, the bridge dynamics is self-similar with a height growing as $t^{2/3}$. This regime extends as long as the initial meniscus can be approximated by a wedge, as shown by this geometric interpretation of coalescence, provided by (\ref{eq:finitesizeeffects}). This asymptotic regime completely disappears at $\theta = 90^\circ$. In this limit the proportionality of $h_0$ and $w$ breaks down since $w \sim h_0^2/R$. \textcolor{red}{This scaling is confirmed by the self-similar collapse presented in [fig. \ref{fig3}(b)] for $\theta=90^{\circ}$, comparing the profiles at 5 different times after contact. Introducing this outer length scale, a new scaling law $h_0 \sim t^{1/2}$ emerges.} 
%Our theory as well as our experiments for $\theta=90^{\circ}$ indeed recovers the coalescence law for freely suspended spherical drops in the inertial regime \cite{Eggers:2003,Aarts:2005, Paulsen:2012}. 
It was found that the exponent of coalescence can be either $2/3$ or $1/2$, depending on the shape at contact -- ``merging wedges'' versus ``merging spheres''. More generally, we expect that the importance of geometry in fast capillary dynamic extends beyond coalescence, as also shown for spreading \cite{Courbin:2009}. In our experiments we have used ultra-pure water, but our results apply whenever inertia is the limiting factor for coalescence. Apart from a short regime where viscous effects cannot be ignored \cite{Paulsen:2012,endnote29}, inertia dominates for aqueous solutions or other low-viscosity fluids. Our results therefore hold for a broad variety of applications such as inkjet printing \cite{Wijshoff:2010}, drop manipulation on a substrate \cite{Karpitschka:2012}, deposition of pesticides on leaves \cite{Vovelle:2000} or cooling and condensation phenomena \cite{Boreyko:2009,Blaschke:2012, Andrieu:2002}.

\acknowledgments{We are grateful to J.F. Hernandez-Sanchez and G. Lajoinie for their help during setting up the experiment and to J. Eggers, D. Lohse and M. Riepen for discussions. This work was funded by NWO through a VIDI grant N$^\circ$11304, and is part of the research program ``Contact Line Control during Wetting and Dewetting'' funded by FOM, ASML and Oc\'{e}.}

\end{document}